\documentclass[aps,10pt,twocolumn,superscriptaddress,floatfix,preprintnumbers,prl,nofootinbib]{revtex4-2}
\usepackage[colorlinks=true,breaklinks=true]{hyperref}
\usepackage[utf8]{inputenc}
\hypersetup{allcolors=[rgb]{0.0 0.0 0.6},linkcolor=[rgb]{0.75 0.05 0.05}}
\usepackage{amsmath,amssymb}
\usepackage{mathtools}
\usepackage[normalem]{ulem}
\usepackage{bm}
\usepackage{float}
\usepackage{physics}
\usepackage{siunitx}
\DeclareSIUnit \parsec {pc}
\DeclareSIUnit \years {yr}
\DeclareSIUnit \electronvolt {eV}
\usepackage[dvipsnames]{xcolor}
\usepackage{float}
\usepackage{graphicx}
\usepackage{epsfig}
\usepackage{letltxmacro}
\LetLtxMacro{\oldcite}{\cite}
\renewcommand{\cite}[1]{\mbox{\oldcite{#1}}}

\usepackage{orcidlink}

\long\def\exclude#1{}

\DeclareSymbolFont{starfontsym}{OT1}{sts}{m}{n}
\DeclareMathSymbol{\mathTerra}{\mathord}{starfontsym}{76}

\newcommand{\beq}{\begin{equation}}
\newcommand{\eeq}{\end{equation}}

\def\ga{\,\,\raise0.14em\hbox{$>$}\kern-0.76em\lower0.28em\hbox
{$\sim$}\,\,}

\newcommand{\wP}{\omega_{\rm pl}}

\long\def\exclude#1{}

\allowdisplaybreaks

\def\bea{\begin{eqnarray}}
\def\eea{\end{eqnarray}}
\def\beq{\begin{equation}}
\def\eeq{\end{equation}}

\hypersetup{
    colorlinks=true,
    citecolor=[rgb]{.1, .7, .6},
    linkcolor=[rgb]{.2, .55, .95},
    filecolor=magenta,
    urlcolor=[rgb]{.1, .7, .6},
}

\begin{document}

\title{Cooling the Shock: New Supernova Constraints on Dark Photons
}

\author{Andrea Caputo\orcidlink{0000-0003-3516-8332}} 
\affiliation{Department of Theoretical Physics, CERN, Esplanade des Particules 1, P.O.~Box 1211, Geneva 23, Switzerland}
\affiliation{Dipartimento di Fisica, ``Sapienza'' Universit\`a di Roma \& Sezione INFN Roma 1, Piazzale Aldo Moro
5, 00185, Roma, Italy}

\author{Hans-Thomas Janka\orcidlink{0000-0002-0831-3330}}
\affiliation{Max-Planck-Institut f\"ur Astrophysik, Karl-Schwarzschild-Str.~1, 
85748 Garching, Germany}

\author{Georg Raffelt\orcidlink{0000-0002-0199-9560}}
\affiliation{Max-Planck-Institut f\"ur Physik, Boltzmannstr.~8, 
85748 Garching, Germany}

\author{Seokhoon Yun\orcidlink{0000-0002-7960-3933}}
\affiliation{Center for Theoretical Physics of the Universe, Institute for Basic Science (IBS), Daejeon, 34126, Korea}

\preprint{CERN-TH-2025-020}
\preprint{CTPU-PTC-25-05}


\begin{abstract}
During the accretion phase of a core-collapse supernova (SN), dark-photon (DP) cooling can be largest in the gain layer below the stalled shock wave. In this way, it could counter-act the usual shock rejuvenation by neutrino energy deposition and thus prevent the explosion. This peculiar energy-loss profile derives from the resonant nature of DP production. The largest cooling and thus strongest constraints obtain for DP masses of \hbox{0.1--0.4\,MeV}, a range corresponding to the photon plasma mass in the gain region. Electron-capture SNe, once observationally unambiguously identified, could provide strong bounds even down to nearly 0.01\,MeV. For a coupling strength so small that neutrino-driven explosions are expected to survive, the DP cooling of the core is too small to modify the neutrino signal, i.e., our new argument supersedes the traditional SN1987A cooling bound.
\end{abstract}

\maketitle

\textbf{\textit{Introduction}}.---Collapsing stars are powerful sources of neutrinos and, conceivably, new feebly interacting particles \cite{Antel:2023hkf}, such as axions, axion-like particles, sterile neutrinos, dark photons, and many others. The widely accepted Bethe-Wilson mechanism \cite{Bethe:1985sox} holds that the associated supernova (SN) explosion is driven by neutrino energy deposition behind the stalled shock wave that has formed at core bounce and so, these elusive particles are actually thought to spawn some of the most dynamical astrophysical class of events \cite{Janka:2012wk, Janka2017Handbooka, Burrows:2020qrp, Boccioli:2024abp}. Moreover, the neutrino signal was observed once from SN1987A and, within sparse statistics, broadly agrees with expectations \cite{Loredo:2001rx, Fiorillo:2023frv, Bozza+2025}. Depending on their exact nature and properties, feebly interacting particles could likewise strongly affect the explosion, proto-neutron star (PNS) cooling, or show up directly through decays into $\gamma$-rays, charged leptons, or neutrinos. These latter effects can pertain to SN1987A, neutron stars (NSs), and the cosmic diffuse flux from all past SNe. The literature on these subjects is vast and has recently grown fast, but there are only partial overall reviews of recent developments \cite{Raffelt:1990yz, Caputo:2024oqc, Carenza:2024ehj, cajohare}.

Dark photons (DPs)~\cite{Fabbrichesi:2020wbt, Caputo:2021eaa}, also called hidden photons, are an intriguing and widely discussed class of particles that could constitute the cosmic dark matter with masses much smaller than traditional WIMPs. They also appear as mediators (a vector portal) to beyond the Standard Model (BSM) physics~\cite{Bjorken:2009mm, Ilten:2018crw}. They kinematically mix with the usual photons and can thus interact with charged particles and there can be $\gamma'\leftrightarrow\gamma$ oscillations. These are resonant if the photon plasma mass $\wP$ equals the DP mass $m_{\gamma'}$, or for the conversion to longitudinal plasmons, if $m_{\gamma'}<\wP$.

There exist many astrophysical constraints on $m_{\gamma'}$ and the mixing parameter $\varepsilon$, defined such that in vacuum, $\gamma'$ sees the effective charge $\varepsilon e$, e.g., on an electron. In particular, SN1987A bounds have often been cited~\cite{Chang:2016ntp, Rrapaj:2015wgs} that rely on the traditional argument that too much PNS cooling during the first second would excessively shorten the observed neutrino signal duration. However, for any particle constraint that relies on a certain phase of stellar evolution, one minimal consistency requirement is that preceding phases are not more strongly affected by the same particles.

The main idea of this \textit{Letter} is to show that this is the case for the SN1987A bound on DPs in that these can efficiently drain energy from the gain region below the stalled bounce shock. The Bethe-Wilson explosion mechanism holds that there is a gain region $r_{\rm gain} < r < r_{\rm shock}$, where the net effect for neutrinos is to deposit energy, thus reviving the shock wave (see also {\em End Material}). The net effect of DPs, on the other hand, is energy drain from the same region, which prevents a sufficient net gain if the new cooling effect is strong enough. The origin for the strong energy loss derives from $\gamma\leftrightarrow\gamma'$ oscillations being resonant for $m_{\gamma'}=0.1$--0.4~MeV, corresponding to $\wP$ in the gain region. In this scenario, any argument based on subsequent PNS cooling would be moot, although to avoid our new effect (SNe by definition explode), one finds more restrictive limits that are moreover independent of SN1987A.

\textbf{\textit{DP model and production channels.}}---The low-energy effective Lagrangian for $A_\mu$ and $A_\mu^\prime$ is~\cite{Holdom:1985ag, Fabbrichesi:2020wbt}
\beq
\begin{split}
\mathcal{L} = & -\frac{1}{4}F_{\mu\nu}F^{\mu\nu} -\frac{1}{4}F^{\prime}_{\mu\nu} F^{\prime \mu\nu}  +\frac{\varepsilon}{2}F_{\mu\nu}F^{\prime \mu\nu} \\[1ex]
&   + \frac{m_{\gamma^\prime}}{2} A_{\mu}^\prime A^{\prime \mu} + eA_\mu J_{\rm EM}^{\mu} \, ,
\end{split}
\eeq
where $F_{\mu\nu}^{(\prime)} = \partial_\mu A_\nu^{(\prime)} - \partial_\nu A_\mu^{(\prime)}$ is the photon (DP) field-strength tensor, $J_{\rm EM}^\mu$ corresponds to the electromagnetic current, and $\varepsilon$ is the kinetic mixing, which sets the interaction strength between the DP and the Standard Model (SM). In vacuum, where the photon is massless, one can diagonalize the kinetic terms and obtain the interaction term $\varepsilon e A_\mu^\prime J_{\rm EM}^\mu$. 

In a dense medium, the effective interactions between DP and the SM are modified by plasma effects and the interaction becomes 
\bea
\varepsilon e A_\mu^\prime J_{\rm EM}^\mu \longrightarrow \varepsilon e \frac{m_{\gamma^\prime}^2}{m_{\gamma^\prime}^2 - \pi_{\rm T,L}} A_\mu^\prime J_{\rm EM}^\mu \,
\label{eq:EffDPcoupling}
\eea
in terms of the transverse and longitudinal projections $\pi_{\rm T, L}$ of the photon polarization tensor. The DP cooling rate per unit volume is then~\cite{Chang:2016ntp, An:2013yfc, Redondo:2013lna}
\bea
\kern-2em{Q}_{\gamma^\prime}
& = &  \frac{\varepsilon^2 m_{\gamma^\prime}^4}{2\pi^2 } \int d\omega \frac{\omega^2 v_{\gamma^\prime}}{e^{\omega/T}-1} \nonumber\\
&&\kern3em{}\times \sum_{i={\rm T,L}} \frac{g_i\left|{\rm Im}\,\pi_i\right|}{\left(m_{\gamma'}^2 - {\rm Re}\,\pi_i\right)^2+\left|{\rm Im}\,\pi_i\right|^2},
\label{eq:SNDPvolumeEm}
\eea
where $v_{\gamma'}=(1-m_{\gamma^\prime}^2/\omega^2)^{1/2}$, $g_{\rm T} = 2$ for the transverse and $g_{\rm L}=1$ for the longitudinal polarization. We have used the fact that the DP coupling structure precisely follows that of ordinary photons after including the extra factors entering the coupling to $J^\mu_{\rm EM}$. 

The dominant contribution to ${\rm Im}\,\pi_i$ and thus to
DP production inside the PNS, is nucleon bremsstrahlung $n+p \rightarrow n + p + \gamma^\prime$ \cite{Rrapaj:2015wgs, Shin:2021bvz}. We use the soft-radiation approximation \cite{Low:1958sn, Nyman:1973gw, Rrapaj:2015wgs} and assume the nuclear medium to be nondegenerate and nonrelativistic \cite{Rrapaj:2015wgs, Chang:2016ntp}. Outside the PNS, the dominant channel is Compton-like scattering, $e^- + \gamma \rightarrow e^- + \gamma^\prime$.

The energy integral in Eq.~\eqref{eq:SNDPvolumeEm} can be split into a region around the resonance and the rest. Because ${\rm Im}\,\pi_i \ll {\rm Re}\,\pi_i$ at the resonance energy $\omega_{\rm res}$, we can approximate
$\bigl[(m_{\gamma^\prime}^2-{\rm Re}\,\pi_i)^2+\left|{\rm Im}\,\pi_i\right|^2\bigr]^{-1} \simeq \bigl[\left(\partial_\omega{\rm Re}\,\pi_i\right)^2\left(\omega - \omega_{\rm res}\right)^2 + \left|{\rm Im}\,\pi_i\right|^2\bigr]^{-1} \simeq \delta (\omega-\omega_{\rm res}) \times \pi\left(\partial_\omega{\rm Re}\,\pi_i\right)^{-1}|{\rm Im}\,\pi_i|^{-1}$.
Therefore, the production rate from the resonance is
\begin{equation}
{Q}_{\gamma'}\big|_{\rm res} \simeq\frac{\varepsilon^2 m_{\gamma^\prime}^4}{2\pi} \sum_{i={\rm T,L}}g_i \left|\frac{\partial {\rm Re}\,\pi_i}{\partial \omega}\right|^{-1} \frac{\omega_{\rm res}^2 v_{\rm res}}{e^{\omega_{\rm res}/T}-1} \,,
\end{equation}
where $v_{\rm res} = (1-m_{\gamma^\prime}^2/\omega_{\rm res}^2)^{1/2}$ and
\begin{equation}
  \left|\frac{\partial {\rm Re}\,\pi_i}{\partial \omega}\right| = \frac{m_{\gamma^\prime}^2}{\omega_{\rm res}v_{\rm res}^2}\left(2+\frac{m_{\gamma^\prime}^2}{\omega_{\rm res}^2}-\frac{3\omega_{\rm pl}^2}{m_{\gamma^\prime}^2}\left[\frac{m_{\gamma^\prime}^2}{\omega_{\rm res}^2}\right]_{i={\rm L}}\right).  
\end{equation}
For longitudinal modes, resonant conversion is always possible for $\omega_{\rm pl} > m_{\gamma'}$, while for the transverse ones, resonance occurs only where $\sqrt{2/3}\, m_{\gamma'} < \omega_{\rm pl} < m_{\gamma'}$. 

\begin{figure*}[ht]
    \centering
      \includegraphics[width=0.9\columnwidth]{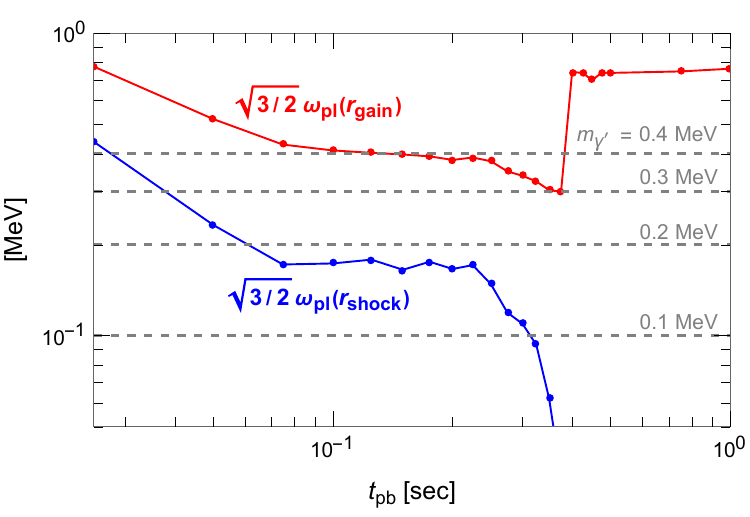}
      \kern2em
      \includegraphics[width=0.9\columnwidth]{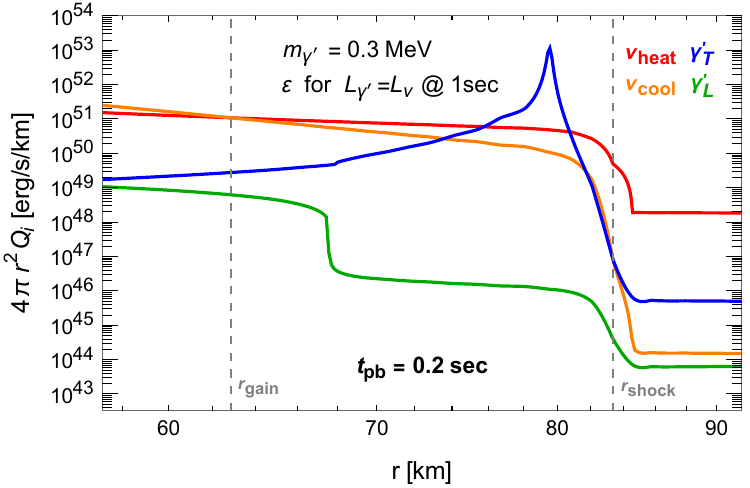}
    \caption{Numerical results for SN explosion model s18.88-LS220, 1D counterpart of 3D models in \cite{Bollig:2020phc}. \textbf{Left:} Time evolution of the plasma frequency at the position of the gain radius (red) and shock radius (blue). When a horizontal dashed line for a given DP mass falls between the red and blue curves, then transverse DPs are resonantly produced in the gain layer. \textbf{Right:} Differential DP luminosity as a function of radius 0.2~s pb for $m_{\gamma'}=0.3\,\rm MeV$ and $\varepsilon \simeq 10^{-8}$, the SN1987A cooling limit. {\em Green~curve} for longitudinal DPs, {\em blue} for transverse modes, showing a resonance near 80~km. The {\em orange and red curves} are neutrino cooling and heating. The vertical dashed lines indicate the gain radius $r_\mathrm{gain}$ and shock radius $r_\mathrm{shock}$.}
    \label{Resonance_Differential}
    \vskip-1pt
\end{figure*}

\textbf{\textit{SN models.}}---To make our study more robust, we consider DP production during the neutrino-heating phase of the stalled SN shock in a set of representative 1D SN models with different PNS masses and nuclear equations of state (EoSs), namely two fairly soft ones, SFHo~\cite{Steiner+2013} and LS220~\cite{Lattimer+1991}, and the stiffer one of Shen~\cite{Shen+1998}. 

Specifically, we use the SFHo-18.8 and SFHo-18.6 models of the Garching group \cite{Bollig:2020xdr, CCSNarchive}, which include muons, and produced NSs with baryonic masses of 1.35\,$M_\odot$ and 1.55\,$M_\odot$, respectively. These models were employed by some of us before to study muon-philic boson emission and low-energy SNe \cite{Bollig:2020xdr, Caputo:2021rux, Caputo:2022mah}. Moreover, we consider model s18.88-LS220 as the 1D counterpart of a self-consistent 3D explosion model~\cite{Bollig:2020phc} with a massive 1.81\,$M_\odot$ NS. Finally, we also analyze an 8.8\,$M_\odot$ electron-capture SN (ECSN) model~\cite{Huedepohl+2010}, which was based on the Shen EoS and left a NS with a baryonic mass of 1.37\,$M_\odot$, but did not include muons and PNS convection. Only this model exploded self-consistently in 1D, whereas the other ones were exploded artificially by reducing the density ahead of the stalled SN shock at a suitable time after bounce. 

This selection of models offers a reasonable variety of postshock density profiles for different kinds of SNe. Since 1D explosion models cannot yield realistic neutrino-heating conditions once the explosion has set in, we investigate net DP cooling in models SFHo-18.8, SFHo-18.6, and s18.88-LS220 only \textit{before} the shock has been re-energized and is on its way to explosion beyond 400--500\,km at typically around 0.4\,s post bounce (pb). 

The ECSN model is an exception. Such explosions develop similarly in 1D and multi-D; multi-D hydrodynamics is here not crucial for the neutrino-driven explosion \cite{Kitaura+2006, Janka+2008}. Therefore, in this case we consider DP cooling even after the onset of the explosion to explore the potentially extended parameter space for future self-consistent simulations. Although ECSNe have not been unambiguously observed (see, e.g., the overview in Ref.~\cite{Jerkstrand+2018}), their existence is predicted by stellar evolution theory~\cite{Langer2012} and their neutrino-driven explosions closely resemble those of low-mass iron-core progenitors with steep outer density profiles, so-called ECSN-like explosions \cite{Melson+2015, Mueller2016, Stockinger+2020}.    

Our analysis demonstrates that different model conditions do not alter our main conclusions, but only mildly change the parameter space of interest, implying that the impact of DP cooling is robust against variations of the SN profile. Next we present results based on model s18.88-LS220, because here the shock behaves similar to the corresponding 3D simulation (though the mass in the gain layer can still be different from the 3D model). Results for the other models are in the {\em End Material}.

\textbf{\textit{Explosion failure for small DP masses.}}---As a first comparison, we show in Fig.~\ref{Resonance_Differential} (left panel) the time evolution of the plasma frequency, rescaled by $\sqrt{3/2}$, at the positions of the gain (red) and shock radius (blue) for all snapshots up to $1\,{\rm s}$ pb. For $m_{\gamma'}\sim 0.2$--$0.4\,\rm MeV$, a resonance crossing occurs for the T modes within the gain layer before the shock accelerates outward at $t \sim 0.4$\,s, potentially jeopardizing the explosion. Such a resonance implies that transverse DP production can efficiently extract energy from this region.

To investigate the impact of this cooling channel, we have computed the differential DP luminosity $dL_{\gamma'}/dr=4\pi r^2 Q_{\gamma'}$ as a function of radius and compared it with neutrino heating and cooling at various times before shock revival. The neutrino heating and cooling terms are (approximately) computed with Eqs.~(28) and~(31) (first expressions) of Ref.~\cite{Janka:2000bt}, which we report in Eqs.~\eqref{Eq:HeatingUsed} and~\eqref{eq:cool} of our \textit{End Material}. Figure~\ref{Resonance_Differential} (right panel) shows an example at 0.2\,s pb, well before shock revival, for $m_{\gamma'}=0.3\,\rm MeV$ and $\varepsilon \simeq 10^{-8}$, the limiting value from the SN1987A cooling constraint. The green curve represents longitudinal DPs, which are here negligible, whereas the blue curve (T modes) reveals a resonance near $r\simeq 80 \, \rm km$. The orange and red curves represent neutrino cooling and heating, respectively, and the gain radius $r_\mathrm{gain}$ is the crossing point, where heating takes over. Evidently, DP cooling exceeds neutrino heating by far in a narrow, but sizable region within the gain layer for the chosen $\varepsilon$ value. For $m_{\gamma'}\lesssim 0.4\,\rm MeV$, DP losses from below the cooling layer ($r < r_\mathrm{gain}$) are negligible.

We have therefore scanned the $m_{\gamma^\prime}$--$\varepsilon$ parameter space and all available SN snapshots, determining when DP cooling integrated over the gain region and a chosen time interval is a fraction $\xi_{\gamma'/\nu}$ of the net neutrino energy deposition in the same region and period. 
For the time integration we used the interval between $t_{\rm st} \sim 70$~ms pb, which is roughly when the shock stagnates at a transiently steady radius, and $t_{\rm sh} \sim 0.4\,{\rm s}$ pb, at which time the shock in the model has expanded to 400--500\,km, initiating the explosion. This is the period when neutrino heating determines the success or failure of the shock reheating mechanism, assisted by strong postshock convection in 3D models. The fact that our investigated 1D models capture the heating conditions only approximately adds into the general uncertainty of the best choice for the critical $\xi_{\gamma'/\nu}$ value.

\begin{figure}[b]
\centering
\includegraphics[width=1.0\columnwidth]{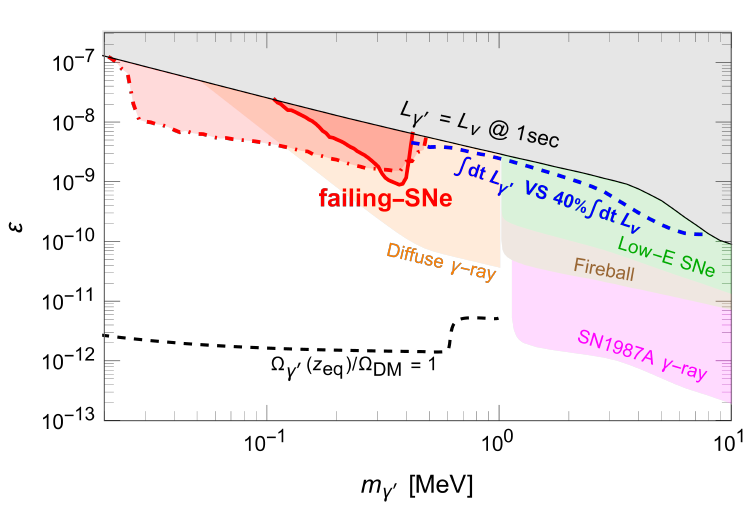}
\caption{SN related DP constraints, all computed in the same SN model; for references see main text. {\em Solid red:} Above this line, DP cooling prevents shock revival by neutrino heating (this work) using $\xi_{\gamma'/\nu} = 0.2$.  {\em Dot-dashed red:} same for the ECSN-like model, but with $\xi_{\gamma'/\nu} = 0.5$. {\em Blue dashed:} DP luminosity dominates before shock revival for comparison. {\em Black line and shaded gray region:} excessive PNS cooling excluded by SN1987A. {\em Orange:} excessive diffuse $\gamma$-ray flux from DP decay emitted by all cosmic SNe. {\em Magenta:} $\gamma$-ray limits from SN1987A. {\em Green:} excessive energy deposition in low-E SNe, causing excessive explosion energies. {\em Brown:} DP-induced fireball formation. {\em Dashed black:} above this line, DPs are thermally overproduced before matter-radiation equality in standard cosmology.} \label{fig:constraints}
\end{figure}

Figure~\ref{fig:constraints} shows the parameters for which DP cooling hinders shock revival. The solid red line marks parameters for which $\xi_{\gamma'/\nu} = 0.2$, although the precise value that prevents explosions can only be determined by self-consistent 3D simulations and depends on the progenitor and somewhat also on SN modeling uncertainties. The value chosen for Fig.~\ref{fig:constraints} should be in the right ballpark, but the final DP limits depend only weakly on this choice because $\varepsilon \propto \xi_{\gamma'/\nu}^{1/2}$. Along this red curve, DP emission from the PNS core is negligible, ensuring self-consistency of the SN model. Above this curve, no explosion is expected, but instead black hole formation. Models with different PNS masses would result in different conditions and explore other regions of parameter space.

For comparison, in Fig.~\ref{fig:constraints} we show analogous results (dot-dashed red line) for our ECSN model; it resembles other ECSN-like explosions as discussed earlier. We chose a slightly larger value of $\xi_{\gamma'/\nu} = 0.5$, because the considered low-mass progenitors explode more easily due to their low binding energies of the matter outside their degenerate cores. Moreover, we integrated up to $t_{\rm sh} \sim 0.5$\,s, after which the explosion energy is saturated, and therefore later neutrino heating is not of great relevance for the explosion. Interestingly, this model reaches to lower $m_{\gamma'}$ values due to a smaller density in the gain layer and thus covers additional DP parameters.

One may wonder whether efficient DP cooling could cause the shock to retreat, potentially moving the DP resonance out of the gain region. However, either the shock lacks sufficient energy to be rejuvenated, or it expands again with fresh energy input by neutrinos, but encounters an efficient DP resonance once more, again halting its progress. We can note examples of this type of shock retreat and expansion due to modulated heating using simulations of the SFHo-18.8 model, where the shock, after being revived within 50\,ms pb, experiences rapid early expansion within 0.2\,s, but subsequently stalls and retreats to less than 150\,km, before re-expanding again. While this behavior is observed in some 1D models, it is rare in multi-D simulations, at least without DP cooling. Nevertheless, it highlights an important point related to our work that we further discuss in the {\em End Material}.

For comparison, we show in Fig.~\ref{fig:constraints} also the standard cooling limit (gray shaded), nominally obtained by imposing $L_{\gamma'} = L_{\nu}$ at 1\,s pb, where $L_{\nu}$ is the total neutrino luminosity of the Garching simulation. We also show the limits (orange shaded) obtained from the diffuse DPs from all past SNe, which create a cosmic background flux analogous to the diffuse SN neutrino background (DSNB) \cite{Ando:2004hc, Beacom:2010kk, Lunardini:2010ab, Mirizzi:2015eza, Kresse:2020nto}, which subsequently decay into three photons~\cite{McDermott:2017qcg, Pospelov:2008jk}. We integrate the DP luminosity over the full simulation duration.

We show the limit from the absence of prompt $\gamma$-rays from SN1987A~\cite{Caputo:2021rux, Jaeckel:2017tud, Hoof:2022xbe, Caputo:ToAppear} (magenta) and from the explosion energy of low-energy-SNe \cite{Caputo:2022mah, Caputo:ToAppear} (green), which are relevant for larger $m_{\gamma'}$ (see below), where DPs decay into  $e^+e^-$ pairs either outside the progenitor or in the progenitor mantle, respectively. We also computed the limits derived from a fire-ball formation and the data from the Pioneer Venus Orbiter (brown), following Ref.~\cite{Diamond:2023scc}. Other probes for DP masses above $1 \, \rm MeV$ exist in the literature \cite{Balaji:2025alr, Diamond:2021ekg, DeRocco:2019njg}, but they are generally less constraining than low-energy SNe or $\gamma$-rays from SN1987A. These bounds will be evaluated in a forthcoming work~\cite{Caputo:ToAppear}.

Finally, we indicate parameters where thermally produced DPs exceed the dark matter abundance at the time of matter-radiation equality within the simplest $\Lambda$CDM scenario~\cite{Redondo:2008ec,Caputo:ToAppear} (above the dashed black curve).

\begin{figure}[t]
\centering
\includegraphics[width=0.99\columnwidth]{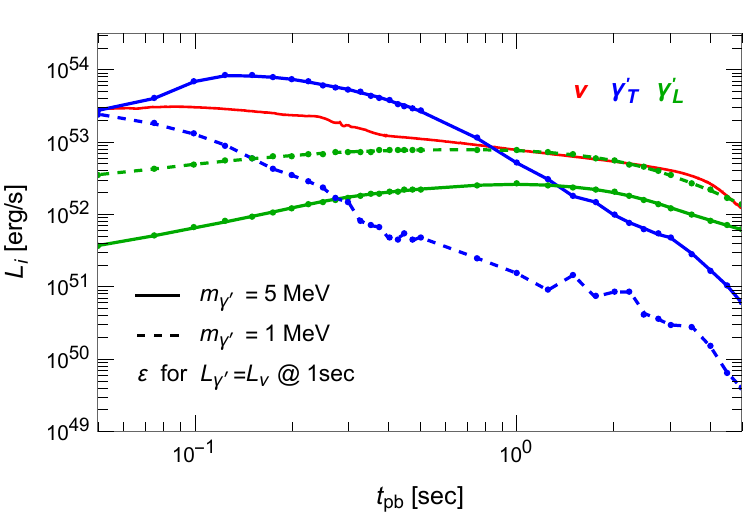}
\caption{DP luminosities for $m_{\gamma'}=5\,{\rm MeV}$ (solid lines) and 1\,MeV (dashed), both for T (blue) and L (green) modes, compared to the total $L_\nu$ (red) from the Garching SN model. For both DP masses, the kinetic mixing has been fixed to match $L_\nu$ at 1\,s, \textit{i.e.,}  $\varepsilon \simeq 3\times 10^{-9}$ and $\varepsilon \simeq 5\times 10^{-10}$ for the DP masses of $1\,{\rm MeV}$ and $5\,{\rm MeV}$, respectively. Contrary to other particle models, where production is highly temperature dependent (e.g., ALPs or QCD axions) and peaks around 1\,s pb, DP production can be more significant at earlier times.}\label{fig:heavyDP}
\end{figure}

\textbf{\textit{More massive DPs.}}---For $m_{\gamma'} \gtrsim 0.5\, \rm MeV$, resonant production of T modes is in deeper regions, below $r_{\rm gain}$. For the first time, we observe that $L_{\gamma'}$ can be comparable to or even exceed $L_\nu$ during very early stages (Fig.~\ref{fig:heavyDP}). Thus one may speculate that DPs might also prevent successful explosions in such cases. Indeed, a \hbox{30--40\%} reduction of the neutrinospheric luminosities could result in a corresponding drop in the energy transfer per nucleon in the gain layer, potentially stalling the shock expansion (see Fig.~6 of Ref.~\cite{Janka:2000bt}). The blue curve in Fig.~\ref{fig:constraints} represents the DP parameters where $L_{\gamma'}$ exceeds 40\% of the neutrinospheric luminosity, integrated up to $t_{\rm sh} \sim 0.4\, \rm s$. In Fig.~\ref{fig:heavyDP}, we also display $L_{\gamma'}$ for masses of $5 \, \rm MeV$ (solid lines) and $1 \, \rm MeV$ (dashed lines), comparing T (blue) and L (green) modes to $L_\nu$ (red). For both DP masses, the kinetic mixing was adjusted to match $L_\nu$ at 1\,s pb. Unlike other particle models, where the production is highly temperature dependent (e.g., ALPs or QCD axions) and peaks around 1\,s pb, DP production is more significant at earlier times. A similar behavior occurs for Majoron production, which primarily depends on the neutrino chemical potential~\cite{Fiorillo:2022cdq}.

However, contrary to DP masses below 0.5~MeV, for more massive DPs the impact on the SN explosion is more subtle, because the particles are produced in the inner regions and therefore the backreaction on the PNS evolution and neutrino emission cannot be neglected. Consequently, the blue curve in Fig.~\ref{fig:constraints} is not a strict bound, but rather an indication of the parameters that might compromise the SN explosion.

\textbf{\textit{Conclusions and future directions.}}---We have highlighted a new physical effect relevant for DPs that was previously overlooked in the SN context. These particles, due to their peculiar resonant production, can be copiously produced within the SN gain layer, potentially preventing the explosion. While most of the relevant DP parameters are already constrained by diffuse $\gamma$-ray data from all SNe, our findings remain significant for several reasons. First, we proved that the usual arguments about PNS cooling are moot, revealing a completely new effect of novel particles on the physics of SN explosions, which may be of interest for other cases as well. Furthermore, this phenomenon could become the leading one in extended dark sectors, where invisible decays exist, suppressing the branching ratios for DP decays into SM particles, thereby weakening the constraints from diffuse $\gamma$-rays. Lastly, our results underscore the importance of perturbative SN probes for DPs, which do not rely on the cooling argument from SN1987A.

Our findings motivate further investigation, particularly self-consistent inclusion of massive DPs in SN simulations, ideally in 3D, to consolidate our bounds based on somewhat arbitrary values of $\xi_{\gamma'/\nu}$ applied in a neutrino heating period before shock runaway. This is especially crucial also for the high-mass region (\hbox{$m_{\gamma'}> 1\,$MeV}), where DPs can significantly impact the explosion mechanism, but post-processed SN models alone are insufficient for rigorous analysis because of extra cooling by DPs in the cooling layer. Moreover, it would be interesting to further explore different SN models that allow for resonant conversion for $m_{\gamma'}\alt 0.1$~MeV, where diffuse $\gamma$-ray limits quickly become irrelevant, given that the DP decay rate scales as $m_{\gamma'}^9$~\cite{McDermott:2017qcg, Pospelov:2008jk}. We leave these studies for future work.

\textbf{\textit{Acknowledgments}}.---We thank Edoardo Vitagliano for useful discussions. AC is supported by an ERC STG grant (``AstroDarkLS'', Grant No.\ 101117510). HTJ and GGR acknowledge partial support by the German Research Foundation (DFG) through the Collaborative Research Centre ``Neutrinos and Dark Matter in Astro- and Particle Physics (NDM),'' Grant SFB-1258-283604770, and under Germany’s Excellence Strategy through the Cluster of Excellence ORIGINS EXC-2094-390783311. SY was supported by IBS under the project code, IBS-R018-D1.
AC thanks the hospitality and support of IBS-CTPU-PTC and support of CKC for the CTPU-CKC Joint Focus Program: Let there be light (particles) Workshop, where part of this work was performed.

\bibliographystyle{bibi}
\bibliography{biblio}

\onecolumngrid
\clearpage

\begin{figure}[ht]
    \centering
\includegraphics[width=0.42\textwidth]{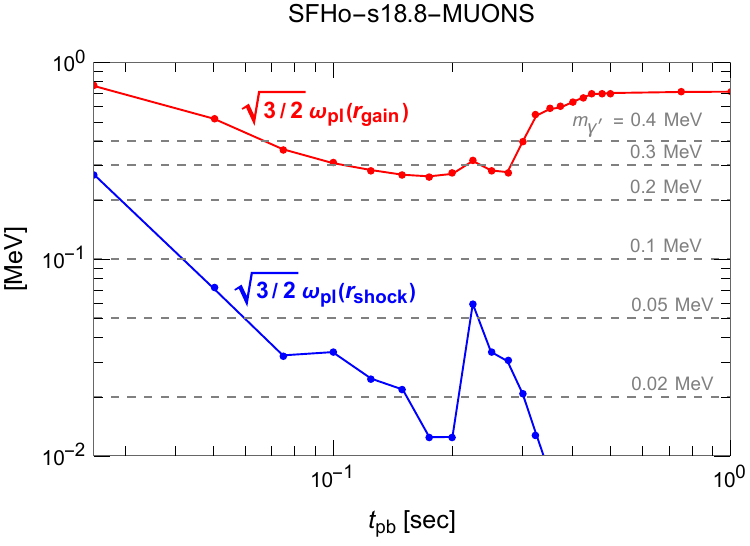}\kern2em
\includegraphics[width=0.42\textwidth]{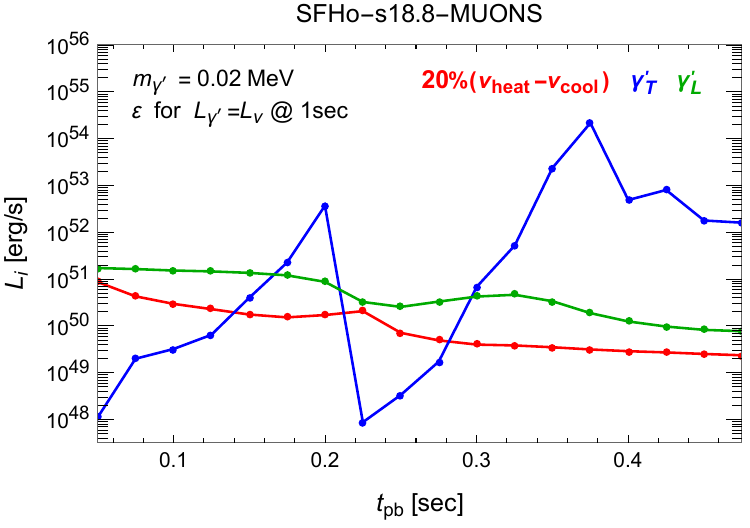}
    \caption{Similar to Fig.~\ref{Resonance_Differential}, now for model SFHo-18.8 with muons~\cite{Bollig:2020xdr,CCSNarchive}.
    {\bf Left.} The shock retreats at $\sim 0.2$\,s to smaller $R$ and correspondingly larger $\omega_{\rm pl}$ before taking off after $\sim$0.35\,s. \textbf{Right:} Total DP cooling in the gain layer as a function of time for both T (blue) and L (green) modes, compared with 20\% of net neutrino heating (red) for $m_{\gamma'}=0.02$\,MeV and $\varepsilon \simeq  1.5\times 10^{-7}$.
    \label{ShockRetreat}
    }
\vskip12pt
    \centering
\includegraphics[width=0.42\columnwidth]{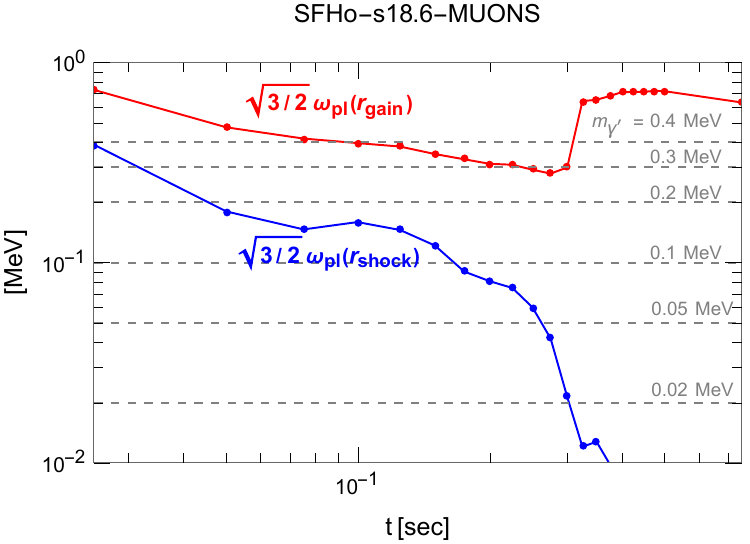}\kern2em
\includegraphics[width=0.42\columnwidth]{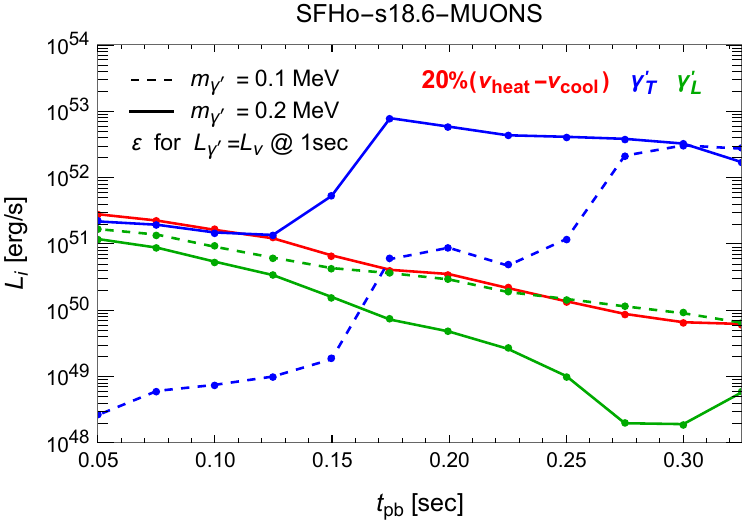}
    \caption{Same as Fig.~\ref{ShockRetreat}, now for model SFHo-18.6 with muons. The shock starts expanding after $\sim$0.3\,s. In the right panel, $\varepsilon \simeq 3\times 10^{-8}$ and $\varepsilon \simeq 1.5\times 10^{-8}$ for the DP masses of $1\,{\rm MeV}$ and $2\,{\rm MeV}$, respectively. 
    \label{SFHo18.6}
    }
\end{figure}

\twocolumngrid

\textbf{\textit{Appendix A: Other simulations.}}---We here present additional simulations. Specifically, Fig.~\ref{ShockRetreat} shows results for Garching model SFHo-18.8 with muons \cite{Bollig:2020xdr,CCSNarchive}. This 1D case is noteworthy because the stalled shock is artificially revived within 50\,ms pb, experiences rapid early expansion within 0.2\,s, and then stagnates. The shock retreats to less than 150\,km before re-expanding again. 

While this behavior is not common for the post-bounce evolution in multi-D simulations, it nevertheless highlights an important point related to our work. When DP cooling becomes significant within the gain region, the shock is expected to lose energy and to retreat, similar to what happened here, albeit, of course, for completely different reasons. This shock retraction causes the gain layer to shift inward, and as a result, for the same DP mass, no resonant production occurs in the gain region during that phase. This behavior is evident in Fig.~\ref{ShockRetreat}. In the right panel, we show the DP cooling rate within the gain region for a representative DP mass of $0.02\,\mathrm{MeV}$. Notably, while the transverse DP luminosity decreases when the shock retreats, two key points emerge. First, the DP luminosity surpasses the neutrino net heating rate (red curve) within the gain region for a significant period ($\sim 0.1\,\mathrm{s}$), ensuring that the integrated effect remains substantial. Second, if DP emission diminishes because the shock retreats, it can re-energize and expand outward again, crossing the resonance once more and ultimately lose energy as predicted. 

In Fig.~\ref{SFHo18.6}, we show analogous results for model SFHo-18.6 with muons, which is a particularly useful case for two main reasons. First, its shock radius is fairly representative, displaying behavior similar to 3D simulations. Second, the PNS mass is smaller, significantly affecting PNS conditions and leading to distinct implications. 

\begin{figure}
    \centering
\includegraphics[width=1.0\columnwidth]{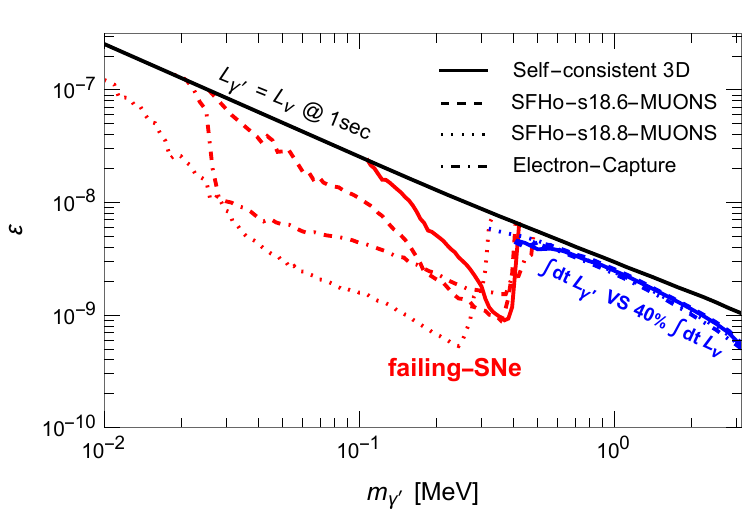}
    \caption{SN related DP constraints from all models. Our new
    ``failing explosion'' criterion is always more stringent than the traditional cooling argument. Moreover, model SFHo-18.6 (dashed) provides constraints very similar to the 1D counterpart s18.88-LS220 (solid red) of the 3D explosion model. SFHo-18.6 (dotted) and the ECSN model (dot-dashed) provide stronger constraints for smaller DP masses because the late-time shock evolution explores lower-density regions. Moreover, the effect of large masses (blue curves) is fairly insensitive to the different models. As stated in the main text, we 
    integrate up to $t_{\rm sh} \simeq 0.5$\,s for the ECSN model, while for all the others, to $t_{\rm sh} \simeq 0.4$\,s.}
    \label{fig:constraints_models}
\end{figure}

In Fig.~\ref{fig:constraints_models}, we finally summarize all constraints. The results of SFHo-18.6 are very similar to those in the main text, primarily due to the comparable shock behavior. Conversely, SFHo-18.8 and the ECSN model yield competitive constraints on much smaller DP masses. This extension derives from the shock now extending to much larger radii, where the density decreases, enabling DP resonant conversion for smaller masses. 

\textbf{\textit{Appendix B: Neutrino heating and cooling in a nutshell.}}---We provide a brief introduction to neutrino heating and cooling in the gain region in the framework of the Bethe-Wilson explosion mechanism. The main processes in this region are charged-current absorption and emission of $\nu_e$ and $\bar{\nu}_e$. Under reasonable approximations, one can derive analytical expressions, useful to develop intuition about their scaling. Assuming, for instance, that $\nu_e$ and $\bar{\nu}_e$ have equal luminosity, $L_{\nu_e} \sim L_{\bar{\nu}_e}$, equal flux factor, and that $0.5\langle \epsilon^2_{\bar{\nu}_e} \rangle \approx \langle \epsilon^2_{\nu_e} \rangle \approx 21 T^2_{\nu_e}$ (neutrino flux spectra close to Fermi-Dirac with no degeneracy), one finds for the heating rate per
unit volume~\cite{Janka:2000bt}
\begin{eqnarray}
Q_{\nu}^{+} &=& \frac{3\alpha^2\!+\!1}{4} \, \frac{\sigma_0 \langle \epsilon_{\nu_e}^2 \rangle}{m_e^2} \, \frac{\rho}{m_u} \, \frac{L_{\nu_e}}{4\pi r^2 \langle \mu_{\nu} \rangle} \, (Y_n + 2Y_p) \label{Eq:HeatingUsed}
\nonumber\\
&\approx& \frac{160~{\rm MeV}}{\rm s} 
\frac{\rho}{m_u} \frac{L_{\nu_e,52}}{r_7^2 \langle \mu_{\nu} \rangle} \left( \frac{T_{\nu_e}}{4 \text{ MeV}} \right)^2 ,  
\label{eq:heat}
\end{eqnarray}
where in the second line we took $Y_n + 2 \, Y_p \approx 1$ for an easier normalization. Here, $\sigma_0 = 1.76 \times 10^{-44} \, \rm cm^2$, $Y_{n,p} = n_{n,p}/n_b$ are the number fractions of free neutrons and protons, $\alpha = -1.26$ is the vector coupling constant, $r_7$ the radius in units of $10^7\,{\rm cm}$, $m_u \simeq 1.66 \times 10^{-24} \,\text{g}$ the atomic mass unit, $L_{\nu_e,52}$ the $\nu_e$ luminosity normalized to $10^{52} \,\text{erg/s}$, and $\langle \mu_{\nu} \rangle$ the neutrino flux factor, which approaches unity in the limit of free streaming, when the neutrinos are forward peaked. The temperature $T_{\nu_e}$ is defined at the neutrinosphere, which, in the context of the gain region, coincides with the energy-sphere—where neutrinos decouple energetically from the background. 

While this does not necessarily align with the sphere of last scattering, the two nearly coincide for $\nu_e$ and $\bar\nu_e$ in SNe. Thus, it is reasonable to refer to {\em the\/} neutrinosphere at radius $r_\nu$, defined by the condition that the effective optical depth satisfies~\cite{rybicki_lightman_1979}:
\begin{equation}
\tau_{\rm eff} = \int^\infty_{r_\nu} dr \, k_{\rm eff}(r) = \frac{2}{3}.
\end{equation}
Here, $k_{\rm eff}$ is the effective opacity, accounting for both scattering and absorption~\cite{Janka:2000bt}. By definition, it follows that $T_{\nu_e} \equiv T(r_\nu)$. Of course, in reality, there is no sharp boundary where neutrinos are fully coupled below and freely streaming above; rather, the transition occurs gradually and is energy dependent.

Along the same lines, the cooling rate can be expressed as~\cite{Janka:2000bt}
\begin{eqnarray}
Q_{\nu}^{-} &=& \left(3\alpha^2\!+\!1\right)  \frac{\sigma_0T^6}{8\pi^2m_e^2}  \frac{\rho}{m_u} \left[Y_p \mathcal{F}_5(\eta_e) + Y_n \mathcal{F}_5(-\eta_e)\right]\label{Eq:CoolingUsed}
\nonumber\\ 
&\approx& \frac{145\,{\rm MeV}}{\rm s} \frac{\rho}{m_u} \left( \frac{T}{2 \text{ MeV}} \right)^6,  \label{eq:cool}
\end{eqnarray}
where $\mathcal{F}_5$ is the Fermi
integral for relativistic particles (Eq.~(32) in Ref.~\cite{Janka:2000bt}), $T$ the local temperature in the gain region, and $\rho$ is the density. In passing to the second line it was assumed that $Y_n + Y_p \approx 1$ and that the electron degeneracy is low because of a large abundance of $e^+e^-$ pairs, which is a valid approximation in shock-heated layers.  An important distinction arises here: while the heating rate in the gain region depends quadratically on the neutrino temperature at the neutrinosphere, the cooling rate scales as the sixth power of the \textit{local} temperature. The gain radius, $r_{\rm gain}$, is then defined as the point where heating and cooling balance, $Q_{\nu}^{+} = Q_{\nu}^{-}$. In units of $10^7 \, \rm cm$, it is given by~\cite{Janka:2000bt}:
\begin{equation}
r_{{\rm gain},7} \left( \frac{T_g}{2 \text{ MeV}} \right)^3 
\approx 1.05 \sqrt{\frac{L_{\nu_e,52}}{\langle \mu_{\nu} \rangle_{\text{g}}}} 
\left( \frac{T_{\nu_e}}{4 \text{ MeV}} \right),
\end{equation}
where $T_g \equiv T(r_{\rm gain})$ is the temperature at that radius.

In the main part of the paper, we use the expressions in the first lines of Eqs.~\eqref{Eq:HeatingUsed} and \eqref{Eq:CoolingUsed} to compute the neutrino heating and cooling rates with the necessary inputs for neutrino and stellar plasma quantities taken from the Garching SN models.

\end{document}